\documentclass{elsart}
\usepackage{psfig}
\usepackage{epsfig}
\usepackage{epsf}
\usepackage{graphics}
\usepackage{amssymb}
\newcommand{\lsim}{\raisebox{-0.07cm}{$\,
\stackrel{<}{{\scriptstyle\sim}}\, $}}

\newcommand{\EPEM}{\mbox{$e^+e^-$}}

\newcommand{\GG}{\mbox{$\gamma\gamma$}}

\newcommand{\GE}{\mbox{$\gamma e$}}

\newcommand{\LGG}{\mbox{$L_{\GG}$}}

\newcommand{\TEV}{\mbox{TeV}}
\newcommand{\WGG}{\mbox{$W_{\gamma\gamma}$}}
\newcommand{\GEV}{\mbox{GeV}}

\newcommand{\CMS}{\mbox{cm$^{-2}$s$^{-1}$}}
\newcommand{\MKM}{\mbox{$\mu$m}}
\providecommand{\NC}{\mbox{${\mathcal NC}$}}
\providecommand{\CC}{\mbox{${\mathcal CC}$}}
\def\ZP #1 #2 #3 {{\it Z.\ Phys.}\ {\bf #1}\ (#2) #3}
\def\PR #1 #2 #3 {{\it Phys.\ Rev.}\ {\bf #1}\ (#2) #3}
\def\SM{$\mathcal{SM}$}
\def\MSSM{$\mathcal{MSSM}$}
\def\2HDM{$2\mathcal{HDM}$}

\providecommand{\g}{\ensuremath{\gamma}}

\providecommand{\W}{\ensuremath{W}}
\providecommand{\Z}{\ensuremath{Z}}

\providecommand{\HPHM}{\ensuremath{H^+H^-}}
\providecommand{\ccbar}{\ensuremath{c\bar c}}
\providecommand{\bbbar}{\ensuremath{b\bar b}}
\providecommand{\qqbar}{\ensuremath{q\bar q}}

\providecommand{\BR}{\mbox{BR}}
\newcommand{\be}{\begin{equation}}
\newcommand{\ee}{\end{equation}}
\newcommand{\bea}{\begin{eqnarray}}
\newcommand{\eea}{\end{eqnarray}}
\newcommand{\bear}{\begin{equation}\begin{array}}
\newcommand{\eear[1]}{\end{array}{#1}\end{equation}}

{\end{list}}
\newcounter{enumct}

\newcommand{\epe}{\mbox{$e^+e^-\,$}}
\newcommand{\ggam}{\mbox{$\gamma\gamma\,$}}
\newcommand{\egam}{\mbox{$e\gamma\,$}}
\newcommand{\gewnu}{\mbox{$e\gamma\to W\nu\,$}}
\newcommand{\eeww}{\mbox{$e^+e^-\to W^+W^-\,$}}
\newcommand{\ggww}{\mbox{$\gamma\gamma\to W^+W^-\,$}}

\begin{document}
\runauthor{Jikia}
\begin{frontmatter}
\title{Gold-plated processes at photon colliders \thanksref{title}}
\author[Moscow]{E.~Boos,}
\author[CERN]{A.~De~Roeck,}
\author[Novosibirsk]{I.F.~Ginzburg,}
\author[KEK]{K.~Hagiwara,}
\author[Hamburg]{R.D.~Heuer,}
\author[Freiburg]{G.~Jikia,\thanksref{Someone}}
\author[Krakow]{J.~Kwiecinski,}
\author[London]{D.J.~Miller,}
\author[Hiroshima]{T.~Takahashi,}
\author[DESY,Budker]{V.I.~Telnov,}
\author[SLAC]{T.~Rizzo,}
\author[Akita]{I.~Watanabe,}
\author[DESY]{P.M.~Zerwas}
\thanks[title]{Talk at the Inter. Workshop on
High Energy Photon Colliders, Hamburg, June 14--17, 2000}
\thanks[Someone]{Corresponding author: jikia@pheno.physik.uni-freiburg.de}
\date{}
\address[Akita]{Akita Keizaihoka University, Akita 010-8515, Japan}
\address[CERN]{CERN, CH-1211 Geneva 23, Switzerland}
\address[DESY]{DESY, Deutsches Elektronen-Synchrotron, D-22603 Hamburg, Germany}
\address[Freiburg]{Universit\"at Freiburg, Hermann--Herder--Str. 3, D-79104 Freiburg, Germany}
\address[Hamburg]{Universit\"at Hamburg/DESY, II Institut f\"ur Experimental Physik, Notkestrasse 85,
D-22607 Hamburg, Germany}
\address[Hiroshima]{Hiroshima University, 1-3-1 Kagamiyama, Higashi-Hiroshima, 739, Japan}
\address[KEK]{Theory Group, KEK, Tsukuba, Ibaraki 305-0801, Japan}
\address[Krakow]{H. Niewodnicza\'nski Institute of Nuclear Physics, Krak\'ow, Poland}
\address[London]{University College London, London WC1E 6BT, UK}
\address[Moscow]{Institute of Nuclear Physics, Moscow State University}
\address[Budker]{Institute of Nuclear Physics, 630090, Novosibirsk, Russia}
\address[Novosibirsk]{Sobolev Institute of Mathematics SB RAS, 630090, Novosibirsk, Russia}
\address[SLAC]{Stanford Linear Accelerator Center, Stanford CA 94309, USA}

\begin{abstract}
We review the most important topics and objectives of the physics
program of the \GG, \GE\ collider (photon collider) option for an
$e^+e^-$ linear collider.
\end{abstract}
\end{frontmatter}

\section{Introduction}

\subsection{Photon Colliders}

A unique feature of an $e^+e^-$ Linear Colliders (LC) with a center of
mass (c.m.s.) energy from a few hundred GeV to several
TeV~\cite{NLC,TESLA,JLC,CLIC} is the possibility to transform it to a
\GG, \GE\ collider (Photon Collider) via the process of Compton
backscattering of laser light off the high energy electrons (positrons
are not needed for photon
colliders)~\cite{GKST81,GKST83,GKST84}. Additional material can be
found in the review
papers~\cite{TEL90,TEL91,BBC,GINSD,TEL92,TEL95,BRODBER,BAIL94,MILLERM,CS,ee97,Tfrei,JLCgg,TESLAgg,Sessler,telnov},
in the Conceptual(Zero) Design Reports~\cite{NLC,TESLA,JLC} and in the
proceedings of the workshop on photon colliders held at
Berkeley~\cite{BERK} in 1995 and in these proceedings~\cite{GG2000}.

The maximum energy of the scattered photons is~\cite{GKST81,GKST83}
\begin{equation}
\omega_m=\frac{x}{x+1}E_0; \;\;\;\;
x \approx \frac{4E_0\omega_0}{m^2c^4}
 \simeq 15.3\left[\frac{E_0}{\TEV}\right]
\left[\frac{\omega_0}{eV}\right],
\end{equation}
where $E_0$ is the electron beam energy and $\omega_0$ the energy of the
laser photon. For example, for $E_0 =250$~GeV, $\omega_0 =1.17$~eV, i.e.
$\lambda=1.06$ \MKM\ (Nd:glass laser), we obtain
$x=4.5$ and $\omega_m = 0.82E_0$.

The high energy photon spectrum becomes more peaked for increasing
values of $x$.  It turns out that the value $x\approx 4.8$ is the
optimum choice for photon colliders, because for $x > 4.8$ the
produced high energy photons create QED \EPEM\ pairs in collision with
the laser photons, and the \GG\ luminosity~\cite{GKST83,TEL90,TEL95}
will be reduced.  Hence, the maximum c.m.s. energy in \GG\ collisions is
about 80\% (and 90\% in \GE\ collisions) of that in \EPEM\
collisions. If smaller photon energies are needed, the same laser can
be used when the electron beam energy is decreased. In this case the
value of the parameter $x$ also decreases and the photon spectrum
becomes less peaked.  Alternatively, a laser with a shorter
wave length may be used  to retain the same sharp spectrum ($x\sim 4.8$)
at lower energy.

A typical luminosity distribution in \GG\ collisions is characterized
by a high energy peak and a low energy part, Ref.\cite{telnov}.  The
peak has a width at half maximum of about 15\%. The photons in the
peak can have a high degree of circular polarization. This peak region
is most useful for experimentation. When comparing event rates in
\GG\ and \EPEM\ collisions we will use the value of the \GG\
luminosity in the peak region $z>0.8z_m$ where
$z=W_{\gamma\gamma}/2E_0$ ($W_{\GG}$ being the \GG\ invariant mass) and
$z_m=\omega_m/E_0$. 
The \GG\ luminosity in this
region is proportional to the geometric luminosity $L_{geom}$ of the electron
beams: $\LGG(z>0.8z_m) \sim 0.1 L_{ee,geom}$.\footnote{for a thickness of
the laser target being equal to one collision length} 
 The geometric luminosity of electron beams in a \GG\
collision region can be made larger than the \EPEM\ luminosity because
beamstrahlung and beam repulsion are absent for photon beams.
  
The luminosity of \GE\ collisions is not proportional to the geometric
electron-electron luminosity (see the figures in Ref.~\cite{telnov})
even in the high energy part of the luminosity spectrum. This is due
to the repulsion of electron beams and beamstrahlung. 

The luminosities expected at the TESLA photon collider~\cite{telnov},
are presented in Table~\ref{tabtel}.

\begin{table}[!hbtp]
\caption{Parameters of a \GG\ collider based on the TESLA design. The
left column refers to $2E_0=500$~GeV, the next two columns are useful for Higgs
studies with $M_h=130$ GeV, for at two different values of $x$.}
\vspace{5mm}
{\renewcommand{\arraystretch}{1.2}
\begin{center}
\begin{tabular}{l c c c} \hline
 & $2E_0=500$ & $2E_0=200$ & $2E_0=158$ \GEV\ \\
 & $x=4.6$ & $x=1.8$ & $x= 4.6$ \\  \hline
$ L_{ee,geom}, [10^{34}$ \CMS] & 12. & 4.8 &  3.8 \\
$ L_{\gamma\gamma} (z>0.8z_{m,\GG\ }),[10^{34} $ \CMS] & 1.15 &
0.35 & 0.36 \\
$ L_{e\gamma} (z>0.8z_{m,\GE\ }),[10^{34}$ \CMS] &
0.97 & 0.31 & 0.27 \\
\end{tabular} \end{center} } \label{tabtel}
\end{table}

For comparison, the ``nominal'' \EPEM\ luminosity at TESLA at
$2E_0=500$~GeV is  $ L_{e^+e^-}(500) = 3\times
10^{34}$ \CMS\ \cite{Brinkmann99}.  For the design parameters of the electron
beams and the same energy, 
\begin{equation}
 L_{\gamma\gamma} (z>0.8z_m) \sim \frac{1}{3} L_{e^+e^-}.
  \label{lgge+e-}
\end{equation}
For beams with even smaller emittances even higher \GG\ luminosities
can be reached, in contrast to the \EPEM\ luminosity which is
restricted by beam collision effects.

The energy spectrum of high energy photons becomes most strongly peaked if the
initial electrons are longitudinally polarized and the laser photons
are circularly polarized.  This gives almost a factor of 3--4 increase
of the luminosity in the high energy peak. The average degree of
circular polarization of photons within the high-energy peak amounts
to $90-95\%$.  The sign of the polarization can easily be changed by
changing the signs of both electron and laser photon initial
polarizations.

Linear polarization of high energy photons can be obtained by using
linearly polarized laser light~\cite{GKST84}. The degree of linear
polarization at maximum energy depends on $x$: $l_{\gamma} =
2(1+x)/(1+(1+x)^2)$, giving 0.334, 0.6, 0.8 for $x=4.8,2,1$
respectively. Polarization asymmetries are proportional to
$l_{\gamma}^2$, therefore low $x$ values are more preferable for this
goal. For $x$ = 2 the maximum energy is only 23\% lower than for
$x=4.8$, but the signal is 3.2 times larger.  So, for energies reduced
by 20\% compared to $x=4.8$, significantly larger linear polarization
can be achieved.

\subsection{Physics Objectives \label{obj}}

Recent experiments at the SLC, LEP, the Tevatron and HERA have
confirmed to high precision the Standard Model (\SM) of
the electroweak interactions. In particular, fermion interactions with
electroweak gauge bosons of the \SM\ were verified to per-mille
precision.  

The central goals of studies at the next generation of \EPEM\ colliders
are the proper understanding of electroweak symmetry breaking, associated
with the problem of mass, and the discovery of new physics beyond the
Standard Model~\cite{peskin,ecfa/desy,white}.  
Three scenarios are possible for future
experiments:
\begin{itemize}
\item
  New particles or interactions will be directly discovered at the
  TEVATRON and LHC. A linear collider  in the \EPEM\ and \GG, \GE\
  modes will then play a crucial role in the detailed and thorough
  study of these new phenomena and in the reconstruction of the
  underlying fundamental theories.  
\item
  LHC and LC will discover and study in detail the Higgs boson but no
  spectacular signatures of new physics or new particles will be
  observed.  In this case the precision studies of potential deviations of
  the properties of the Higgs boson, electroweak gauge bosons and the
  top quark from their Standard Model  predictions can provide
  clues to the physics beyond the \SM.  
\item
  Electroweak symmetry breaking (EWSB) is a dynamical phenomenon. The
  interactions of $W$ bosons and $t$ quarks must then be studied at
  high energies to explore new strong interactions at the \TEV\ scale.
\end{itemize}

Electroweak symmetry breaking in the \SM\ is based on the Higgs
mechanism, which introduces one elementary Higgs boson.  The model
agrees with the present data, and the recent global analysis of
precision electroweak data in the framework of the \SM\ \cite{osaka}
suggests a Higgs boson lighter than $200$ \GEV. A Higgs boson in this
mass range is expected to be discovered at the TEVATRON or the
LHC. However, it will be the LC in all its modes that tests whether
this particle is indeed the \SM\ Higgs boson or whether it is
eventually one of the Higgs states in extended models like the
two--Higgs doublet model (\2HDM), or the minimal supersymmetric
generalization of the \SM, \MSSM . At least five Higgs bosons are
predicted in supersymmetric models, $h^0, H^0, A^0, H^+, H^-$. Unique
opportunities are offered by the photon collider to search for the
heavy Higgs bosons in areas of SUSY parameter space not accessible
elsewhere.

In principle, EWSB could also be generated by a strong-coupling
theory. In such models, the signals of the EWSB mechanism are most
clearly manifest in the properties of the heaviest \SM\ particles, the
$W$ and $Z$ bosons and the $t$-quark. In that case, measurement of the
anomalous electroweak gauge boson and top quark coupling 
at the LC will be among the most central issues of the physics
program.

\begin{figure}[htb]
\begin{center}
\epsfig{file=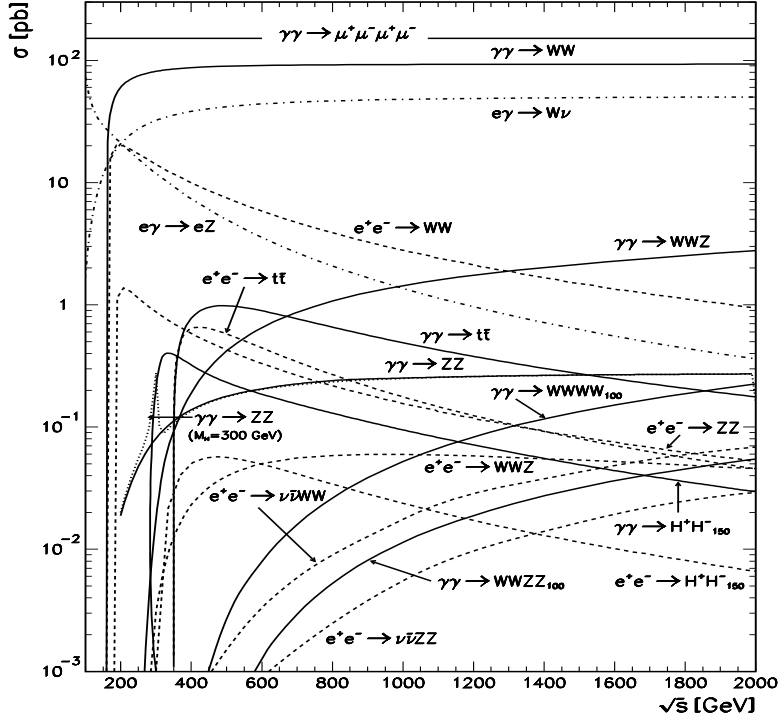,width=0.8\textwidth,height=0.8\textwidth}
\end{center}
\caption{ Typical cross sections in $\gamma\gamma$, $\gamma e$ and
$e^+e^-$ collisions. The polarization is assumed to be zero. Solid,
dash-dotted and dashed curves correspond to $\gamma\gamma$, $\gamma e$
and $e^+e^-$ modes respectively. Unless indicated otherwise the
neutral Higgs mass was taken to be 100~GeV. For charged Higgs pair
production, $M_{H^\pm}=150$~GeV was assumed.}
\label{fig:cs}
\end{figure}

Photon colliders have distinct advantages in searches for and
measurements of new physics objects. In general, phenomena in \EPEM\ and \GG,
\GE\ collisions are quite similar because the same particles can be
produced. However, the reactions are different and often give
complementary information. Some phenomena can be  studied better at
photon colliders due to higher statistical accuracy (based on much larger
cross-sections) or due to higher accessible masses (single resonances
in \GG\ and \GE\ or a pair of light and heavy particles in \GE).
A comparison of  cross-sections for some processes in \EPEM\ and \GG, \GE\ 
collisions are presented in Fig.\ref{fig:cs}~\cite{CS}.

\begin{figure}[!thb]
\centering
\vspace*{-0.9cm}
\hspace*{-0.5cm} \epsfig{file=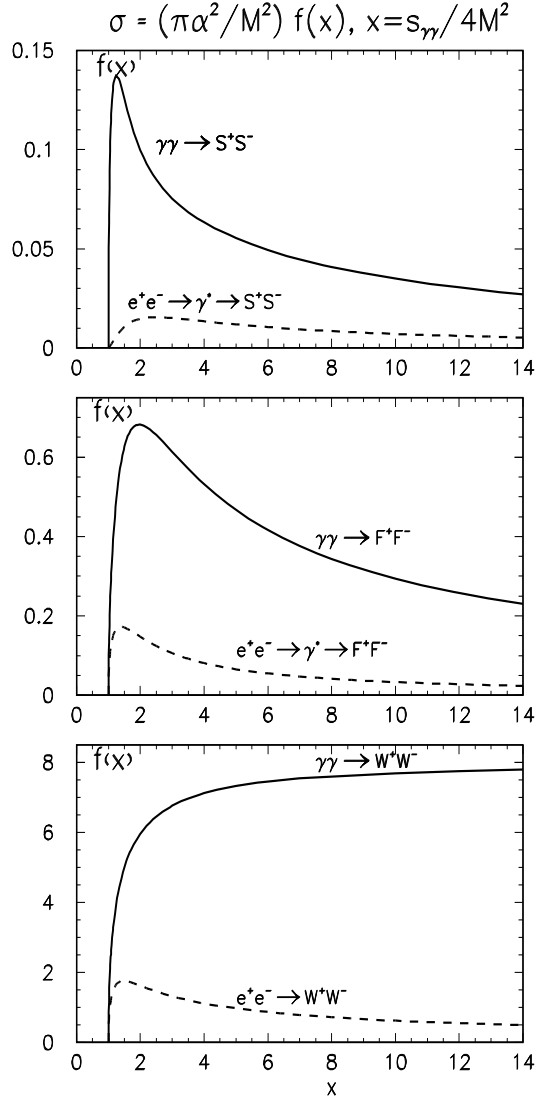,height=16.5cm,width=8.9cm}
\vspace*{-1.cm}
\caption{Comparison between cross sections for charged pair production
in unpolarized \EPEM\ and \GG\ collisions. S (scalars), F (fermions),
W ($W$ bosons); $\sqrt{s}$ is the invariant mass (c.m.s. energy of
colliding beams). The contribution of the \Z\ boson to the production
of S and F in \EPEM\ collisions was not taken into account, it is less
than 10\%.}
\label{charged}
\end{figure}

The cross sections for pairs of scalars, fermions or vectors particles
are all significantly larger (about one order of magnitude) in $\gamma
\gamma$ collisions than in $e^+e^-$ collisions, as demonstrated in
Fig.\ref{charged}~\cite{GSwinter,TEL90,GINSD,TEL95}.

\begin{figure}[!thb]
\centering
\vspace*{-0.7cm}
\hspace*{-0.5cm} \epsfig{file=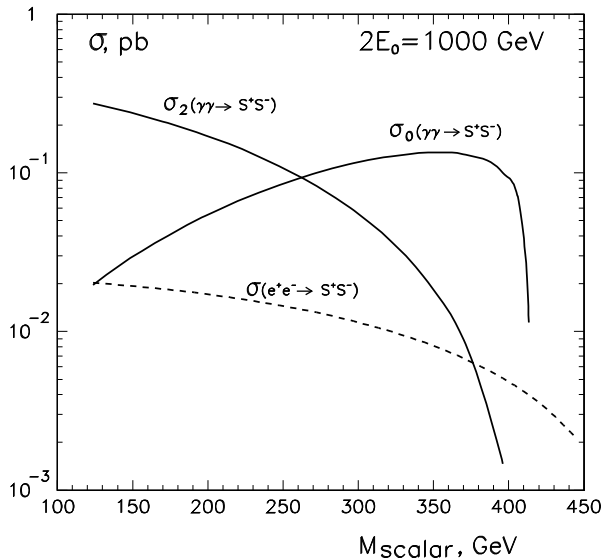,width=11.cm}
\vspace*{-0.7cm}
\caption{The pair production cross section for charged scalars 
  in \EPEM\ and \GG\ collisions at $2E_0$ = 1 TeV collider energy (in \GG\ 
  collision $W_{max}\approx 0.82$ GeV ($x=4.6$)); $\sigma_0$ and
  $\sigma_2$ correspond to the total \GG\ helicity 0 and 2, respectively.}
\vspace{3mm}
\label{crossel}
\end{figure} 

 For example, the maximum cross section for \HPHM\ production with
unpolarized photons is about 7 times larger in \GG\ collisions than
in \EPEM\ collisions (see Fig.\ref{fig:cs}). With polarized photons
and not far from threshold, it is even larger by a factor of 20,
Fig.~\ref{crossel}~\cite{Tfrei}. As a result, for the luminosity given
in the Table~\ref{tabtel} the event rate is 7 times higher.

The two-photon production of pairs of charged particles is a pure QED
process, while the cross section of pair production in \epe\ collision
depends also on the weak isospin of the produced particles via $Z$
exchange, and (sometimes) t--channel exchanges contribute.  Therefore,
measurements of pair production both in \EPEM\ and \GG\ collisions can
be exploited to disentangle various couplings of the charged Higgs
particles.

\begin{figure}[!thb]
  \centering
\vspace*{-0.5cm}
  \hspace*{-0.5cm} \epsfig{file=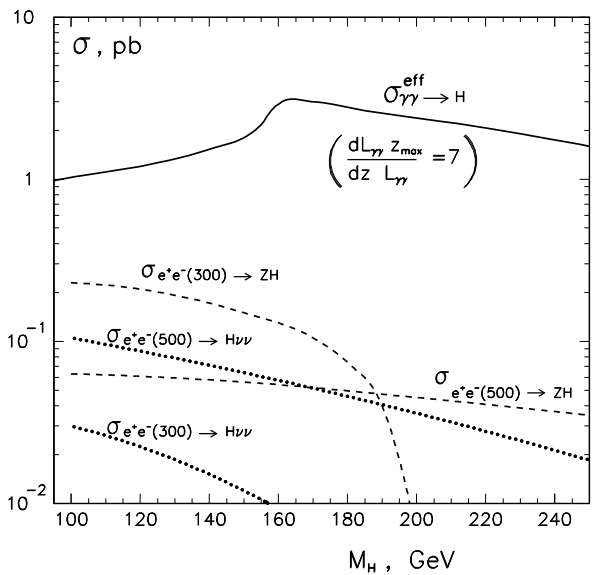,width=11.cm}
\vspace*{-0.7cm}
  \caption{Total cross sections of the Higgs boson production in
    \GG\ and \EPEM\ collisions. To obtain the Higgs boson
    production rate at the photon collider, the cross section should be
    multiplied by the luminosity in the high energy peak
    $L_{\GG}(z>0.65)$ given in Table~\ref{tabtel}.}
  \vspace{3mm}
  \label{hcross}
\end{figure} 

Another example is the direct resonant production of the Higgs boson
in \GG\ collisions. It is evident from Fig.~\ref{hcross}~\cite{ee97},
that the cross section at the photon collider is several times larger
than the Higgs production cross section in \EPEM\ collisions. Although
the \GG\ luminosity is smaller than the \EPEM\ luminosity, the production rate
 of the \SM\ Higgs boson with mass between $100$ and $200$ 
\GEV\ in \GG\ collisions
is nevertheless 1--5 times larger than the rate in \EPEM\ collisions
at $2E_0 = 500$ \GEV.

Photon colliders in the \GE\ mode can produce particles which are
kinematically not accessible at the same collider in the \EPEM\ mode.
For example, in \GE\ collisions a heavy charged particle in
association with a light neutral particle can be produced, such as a
supersymmetric slepton plus a neutralino or a new $W^{\prime}$ boson and
neutrino, and the discovery limits can be extended.

Based on these arguments alone, and without relying on the dynamics
 {\it a priori,} $e^+e^-$ and $\gamma \gamma$/\GE\ modes are expected
 to be nicely complimentary for new physics searches.  Even though the
 analyses of new physics scenarios are not yet as advanced in $\gamma
 \gamma$ and \egam\ collisions as they are for $e^+e^-$ collisions,
 advantages are obvious for some scenarios.

We present a short summary of the most important objectives and topics
of the physics program at the photon collider mode of an LC. We
discuss studies of  Higgs physics (Section~2), supersymmetry
(Section~3), studies of the dynamics of $W$-bosons (Section~4), extra
dimensions (Section~5), top quark physics (Section~6), and QCD and
hadron physics (Section~7). In concluding we present a short list of
processes which appear most important for the physics studies at a
photon collider.

\section{Study of the Higgs Boson}

  The study of the Higgs boson plays an essential role in exploring
the electroweak symmetry breaking and the origin of mass.  The lower
bound on $M_h$ from direct searches at LEP is presently $113.5$ \GEV\
at $95\%$~confidence level (CL)~\cite{igokemenes}. A surplus of events
at LEP provides tantalizing indications of a Higgs boson with $M_h
=115^{+1.3}_{-0.7}$ \GEV\ (90\%~CL)~\cite{higgs115,igokemenes}.
Recent global analyses of precision electroweak data~\cite{osaka}
suggest that the Higgs boson is light, yielding at 95\%~CL that
$M_h=62^{+53}_{-30}$ \GEV.  There is remarkable agreement with the
well known upper bound of $\sim$ 130 \GEV\ for the lightest Higgs
boson mass in the minimal version of supersymmetric
theories~\cite{MSSM}. Such a Higgs boson should definitely be
discovered at the LHC if not already at the TEVATRON.

Once the Higgs boson is discovered, it will be crucial 
to determine the mass, the total width, spin, parity, CP--nature and the
tree--level and one--loop induced couplings in a model--independent way.
Here the \EPEM\ and \GG\ modes of the LC will play a central role.
The \GG\ collider option of an LC offers the unique possibility
to produce the Higgs boson as an s--channel
resonance~\cite{Barklow,GunionHaber,BBC1}:
$$
\GG\to h_0 \to b\bar{b} ,WW^*, ZZ, \tau \tau, gg, \GG\ 
\ldots\,.
$$
 The total width of the Higgs boson at masses below 400 GeV is much
smaller than the characteristic width of the \GG\ luminosity spectra
(FWHM $\sim$ 10--15\%), so that the Higgs production rate is
proportional to $d\LGG/dW_{\GG}$:
\begin{equation}
\dot{N}_{\GG\ \to h}
=\LGG \times \frac{d\LGG M_h}{d\WGG \LGG}\frac{4\pi^2\Gamma_{\GG}
(1+\lambda_1 \lambda_2)}{M^3_h}\equiv \LGG \times \sigma^{eff}.
\end{equation}
where $\lambda_i$ are the photon helicities.  The Higgs search and
study can be done best by exploiting the high energy peak of the \GG\
luminosity spectrum where $d\LGG/d\WGG$ has a maximum and the
photons have a high degree of circular polarization.  The effective
cross section for $(d\LGG/d\WGG) (M_h/\LGG)=7$ and $1+\lambda_1
\lambda_2=2$ is presented in Fig.~\ref{hcross}.  The luminosity in the
high energy luminosity peak ($z>0.8z_m$) was defined in Section 1.
For the luminosities given in Table~\ref{tabtel}, the ratio of the
Higgs rates in \GG\ and \EPEM\ collisions is about 1 to 5 for
$M_h =100$--$200$ \GEV.

The Higgs boson at photon colliders can be detected as a peak in the
invariant mass distribution or (and) can be searched for by energy
scanning using the sharp high energy edge of the luminosity
distribution~\cite{ee97,ohgaki}. The scanning allows also to control
backgrounds. A cut on the acollinearity angle between
two jets from the Higgs decay ($bb,\tau\tau$, for instance) allows to select
events with a narrow distribution (FWHM $\sim$ 8\%)   on the invariant
mass~\cite{TKEK,Tfrei}.

The \GG\ partial width $\Gamma(h\to\GG)$ of the Higgs boson is of
special interest, since it is generated at the one--loop level
including all heavy charged particles with masses generated by the
Higgs mechanism. In this case the heavy particles do not decouple. As
a result, the Higgs cross section in \GG\ collisions is sensitive to
contributions of such particles with masses beyond the energy covered
directly by the accelerators.

Due to the high cross section some Higgs branching ratios can be measured in
\GG\ collisions with accuracies comparable or even better than those
in \EPEM\ collisions. Combined measurements of $\Gamma(h\to\ggam)$
and $\BR(h\to\ggam)$ at the \epe\ and \ggam\ LC provide a model
independent measurement of the total Higgs width~\cite{snow96}.

The required accuracy of the $\Gamma(h\to\GG)$ measurements in the
SUSY sector can be inferred from the results of the studies of the
coupling of the lightest SUSY Higgs boson to two photons in the
decoupling regime~\cite{decoupling}. It was shown that in the
decoupling limit, where all other Higgs bosons and the supersymmetric
particles are very heavy, chargino and top squark loops can generate a
sizable difference between the standard and the SUSY two--photon Higgs
couplings. Typical deviations are at the level of a few percent. Top
squarks heavier than $250$ \GEV\ can induce deviations larger than
$\sim 10\%$ if their couplings to the Higgs boson are large.

The  control of the polarizations of back-scattered photons
provides a powerful tool for exploring the $CP$ properties of any
single neutral Higgs boson that can be produced with reasonable rate
at the photon  collider~\cite{GF92,GK94,KKSZ94}.  $CP$-even
Higgs bosons $h^0$, $H^0$ couple to the combination
$\vec{\varepsilon_1}\cdot \vec{\varepsilon_2}$,
while the $CP$-odd Higgs boson $A^0$ couples to
[$\vec{\varepsilon}_1\times \vec{\varepsilon}_2]\cdot\vec{k_\gamma}$,
where $\vec{\varepsilon}_i$ are photon polarization vectors.  The
scalar Higgs boson couples to linearly polarized photons with a
maximal strength if the polarization vectors are parallel, the
pseudoscalar Higgs boson if the polarization vectors are
perpendicular:
\begin{equation}
\sigma \propto 1 \pm l_{\gamma 1}l_{\gamma 2}\cos{2\phi},
\label{clin}
\end{equation}
$ l_{\gamma i}$ are the degrees of linear polarization and $\phi$ is
the angle between $ \vec{l}_{\gamma 1}$ and $ \vec{l}_{\gamma 2}$. The
signs $\pm$ correspond to the CP$ =\pm 1$ scalar particles.

\begin{figure}[htb]
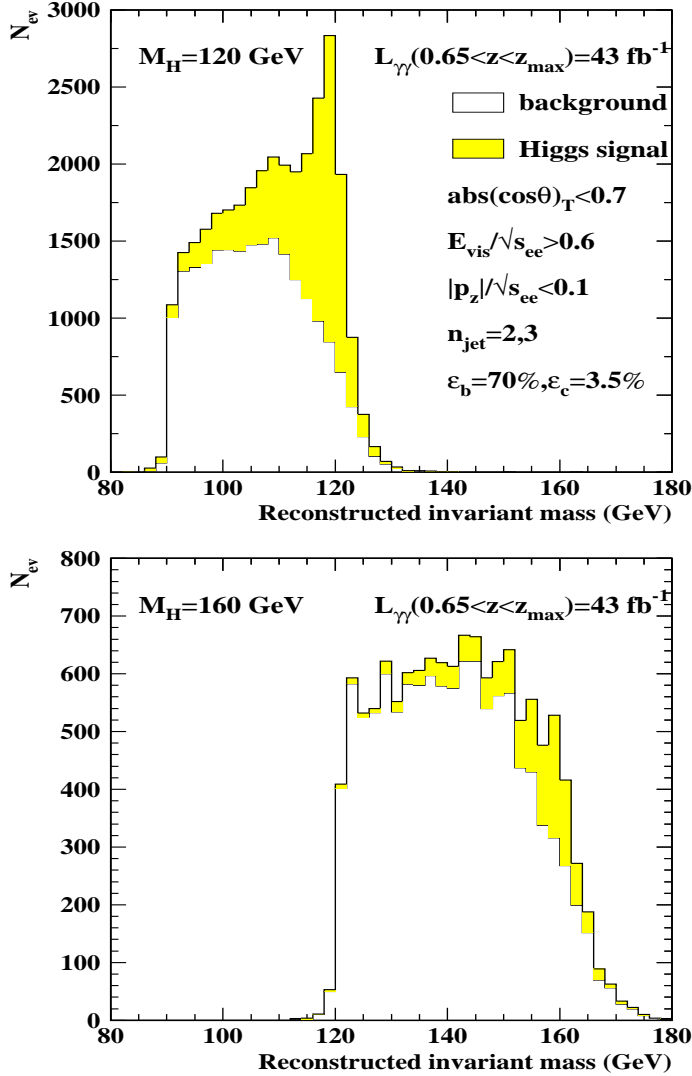

\begin{center}
\epsfig{file=fig5_120.epsi,width=0.65\textwidth,height=0.5\textwidth}

\vspace*{.3cm}

\epsfig{file=fig5_160.epsi,width=0.65\textwidth,height=0.5\textwidth}
\end{center}
\caption{Mass distributions for the Higgs signal and heavy quark
background for a) $M_{\rm h}=120$~GeV and b) $160$~GeV
\cite{Soldner}.}
\label{fig:higgs}
\end{figure}

\subsection{Light \SM\ or \MSSM\ Higgs Boson}

A light Higgs boson $h$ with mass below the $WW$ threshold can be
detected in the $\bbbar$ decay mode. Simulations of this process have
been performed in
Refs.~\cite{BBC1,Ohgaki1997,JLCgg,Ohgaki1998,Melles,Soldner}. The main
background to the $h\to b\bar{b}$ is the continuum production of
$\bbbar$ and $\ccbar$ pairs. A high degree of circular polarization of
the photon beams is crucial in this case, since for equal photon
helicities $(\pm\pm)$, which relevant for the spin--zero resonant states,
the $\GG\to \qqbar$ QED Born cross section is suppressed by a factor
$m^2_q/W_{\GG}^2$~\cite{Barklow,Ispirian}.  Another potentially
dangerous background originates from heavy quark pair production
accompanied by the radiation of an additional gluon, which is not
suppressed even for the equal photon
helicities~\cite{JikiaTkabladze,Khhiggs}.  In addition, virtual QCD
corrections for $J_z=0$ were found to be especially large due to a
double-logarithmic enhancement factor, so that the corrections are
comparable or even larger than the Born contribution for the two-jet
final topologies~\cite{JikiaTkabladze}. Recent studies on Higgs
production at photon colliders~\cite{Soldner,JikiaSoldner,Melles}
include gluon emission in $\gamma\gamma\to \qqbar$ and all
next-to-leading QCD corrections, as well as resummed leading
double-logarithmic corrections~\cite{Sudakov,melles-dl}.

A Monte Carlo simulation of $\GG\to h\to \bbbar$ for $M_h = 120$ and
$160$ \GEV\ has been performed for an integrated luminosity in the
high energy peak of $\LGG(0.8z_m<z<z_m)=43$ fb$^{-1}$ in Ref.
\cite{Soldner,JikiaSoldner}. Real and virtual gluon corrections for
the Higgs signal and the
backgrounds~\cite{JikiaTkabladze,Khhiggs,melles-dl,JikiaSoldner,Melles}
have been taken into account.

The results for the invariant mass distributions for the combined
$\bbbar (\gamma)$ and $\ccbar (\gamma)$ backgrounds, after cuts, and for the
Higgs signal are shown in in Fig.~\ref{fig:higgs}~\cite{Soldner}.  Due
to the large charm production cross-section in $\gamma\gamma$
collisions, excellent $b$ tagging is required~\cite{Soldner,Melles}. A
$b$ tagging efficiency of $70\%$ for $\bar b b$ events and residual
efficiency of $3.5\%$ for $\bar c c$ events were used in these
studies.


For a \ggam\ luminosity in the high energy peak of 43 fb$^{-1}$ a
relative statistical error of  
\be \frac{\Delta[\Gamma(h\to\GG)\BR(h\to
\bbbar)]}{[\Gamma(h\to\GG)\BR(h\to\bbbar)]}\approx 2\%
\label{ggamh}
\ee can be achieved in the Higgs mass range between 120 and 140~GeV.

The statistical accuracy decreases for $M_h>140$~GeV where the $WW^*$
decay mode becomes important and $b\bar b$ is not the dominant decay
mode any more. For $M_h=160$ GeV the statistical error is  about 8\%.

Assuming that the $h\to \bbbar$ branching ratio $\BR(h\to b\bar b$)
can be measured at the LC in \epe\ (and $\gamma\gamma$) collisions
with an accuracy of 1\%~\cite{battaglia} the partial two-photon Higgs
width can be calculated using the relation \be
\Gamma(h\to\GG)=\frac{[\Gamma(h\to\GG) \BR
(h\to\bbbar)]}{[\BR(h\to\bbbar)]} \ee with almost the same accuracy as
in Eq.(\ref{ggamh}).  Such a high precision of the $\Gamma(h\to\GG)$
measurement can only be achieved at the \ggam\ mode of the LC. With
such an accuracy it will be possible to discriminate
between the \SM\ Higgs particle and the lightest scalar Higgs boson of
the \MSSM\ or \2HDM\ \cite{decoupling} and to isolate contributions of
new heavy particles.

The \SM\ Higgs boson with mass $135<M_h<190$~GeV will predominantly
decay into $WW^*$ or $WW$ pairs. This decay mode should permit the
detection of the Higgs signal below and slightly above the threshold
of $WW$ pair production~\cite{WW*}.  In order to determine the
two-photon Higgs width in this case  the relation 
\be
\Gamma(h\to\GG)=\frac{[\Gamma(h\to\GG) \BR (h\to WW^*)]}{[\BR(h\to
  WW^*)]}, 
\ee 
can be used, where $\BR(WW^*)$ is obtained from the measurements of
$\sigma(e^+e^-\to hZ)\times \BR(WW^*)$ and $\sigma(e^+e^-\to hZ)$.
For $M_h= 160$ GeV the  product 
$\Gamma(h\to\GG) \BR (h\to WW^*)$ can be measured at the photon collider
with the statistical accuracy better than 2\% for an integrated \GG\ luminosity
of 40 fb$^{-1}$ in the high energy peak. The accuracy of
$\Gamma(h\to\GG)$ will be determined by the accuracy of $\BR (h\to WW^*)$
in \EPEM\ experiments,  expected to be about 2\%.

Above the $ZZ$ threshold, the most promising channel to detect the
Higgs signal is the reaction $\GG\to ZZ$~\cite{ZZ}.  In order to
suppress the huge background from the tree level $W^+W^-$ pair
production leptonic ($l^+l^-\ l^+l^-$, $BR=1\%$) or semileptonic
($l^+l^-\ q\bar q$, $BR=14\%$) decay modes of the $ZZ$ pairs must be
selected. Although there is only a one-loop induced continuum
production of $ZZ$ pairs in the \SM, a large irreducible background to
the Higgs signal well above the $WW$ threshold is generated in the
continuum~\cite{ZZ}. Due to this background the intermediate mass
Higgs boson signal can be observed at the \GG\  collider in the $ZZ$
mode only if the Higgs mass is less than 350--400~GeV.

Hence, the two-photon \SM\ Higgs width can be measured at the photon
collider, either in $b\bar b$, $WW(WW^*)$ or $ZZ(ZZ^*)$ decay modes, up to the
Higgs mass of 350--400~GeV.  Other decay modes, like $h\to \tau\tau, 
\GG$, can also be explored at photon colliders, but no studies have
been carried out so far.

Assuming that in addition to the measurement of the $h\to \bbbar$
branching ratio also the \BR($h\to\ggam$) can be measured
at the $e^+e^-$ linear collider with an accuracy~\cite{brhgg}
10--15\%, the total width
of the Higgs boson can be determined in a model independent way \be
\Gamma_{h}=\frac{[\Gamma(h\to\GG) \BR (h\to\bbbar)]}{[\BR(h\to\GG)]
  [\BR(h\to\bbbar)]} \ee to an accuracy dominated by the expected
error on $\BR(h\to\gamma\gamma)$. The measurement of this branching ratio
at the photon collider can improve the accuracy of the total Higgs width.

The total Higgs width can also be determined in the $e^+e^-$
collisions using the reaction $\EPEM\to h\nu\bar{\nu}$, in which the
Higgs boson is produced in collisions of virtual \W\ bosons, $\sigma \propto
\Gamma(h\to WW^*)$. Combined with the measurement of $\BR(WW^*)$ in
$e^+e^-$ collisions the total Higgs width can be determined in this
way~\cite{Borisov-Richard,desch-meyer}. The typical precision of such
a measurement for a Higgs mass range of $110$--$140$~GeV is about $(10-3)\%$.

\subsection{Heavy \MSSM\ Higgs Bosons}

The minimal supersymmetric extension of the Standard Model includes
two charged ($H^\pm$) Higgs bosons and three neutral Higgs bosons: the
light $CP$-even Higgs particle ($h$), the heavy $CP$-even ($H$) and
the $CP$-odd ($A$) Higgs states.  For large value of the $A$ mass, the
properties of the light $CP$-even Higgs boson $h$ are similar to the
light \SM\ Higgs boson, and it can be detected in the $b\overline{b}$
decay mode, just as the \SM\ Higgs boson. Its mass is limited to
$M_h\lsim 130$~GeV.  However, the masses of the heavy Higgs bosons
$H,A,H^\pm$ are expected to be of the order of the electroweak scale
up to about one TeV.  They are nearly degenerate. The $WW$ and $ZZ$
decay modes are suppressed for $H$, and these decays are forbidden for
$A$.  Instead of the $WW$, $ZZ$ decay modes, the $t\overline{t}$ decay
channel may be useful, if the Higgs boson masses are heavier than
$2M_t$ and if $\tan\beta \ll 10$.  An important property of the SUSY
couplings is the enhancement of the bottom Yukawa couplings with
increasing $\tan\beta$.  For moderate and large values of $\tan\beta$,
the decay mode to $b\overline{b}$ and to $\tau^+\tau^-$ is
substantial~\cite{Muhlleitner,4b}.

Extensive studies have demonstrated that, while the light Higgs boson
$h$ of the \MSSM\ can be found at the LHC, the heavy bosons $H$ and
$A$ may escape the discovery for intermediate values of $\tan
\beta$~\cite{ATLAS}.  At an $e^+e^-$ LC heavy \MSSM\ Higgs bosons can
only be found in associated production, $e^+e^- \to HA$~\cite{3b},
with $H$ and $A$ having almost equal masses. In the first phase of
the LC collider with a total $e^+e^-$ energy of 500 GeV, the heavy
Higgs bosons can thus be discovered for a mass up to about 250 GeV. To
extend the mass reach by a factor of 1.6, the $\gamma\gamma$ option of
LC can be used, where these bosons can be produced singly.

\begin{figure}[hbt]
\begin{center}
\vspace*{0.2cm}
\epsfig{file=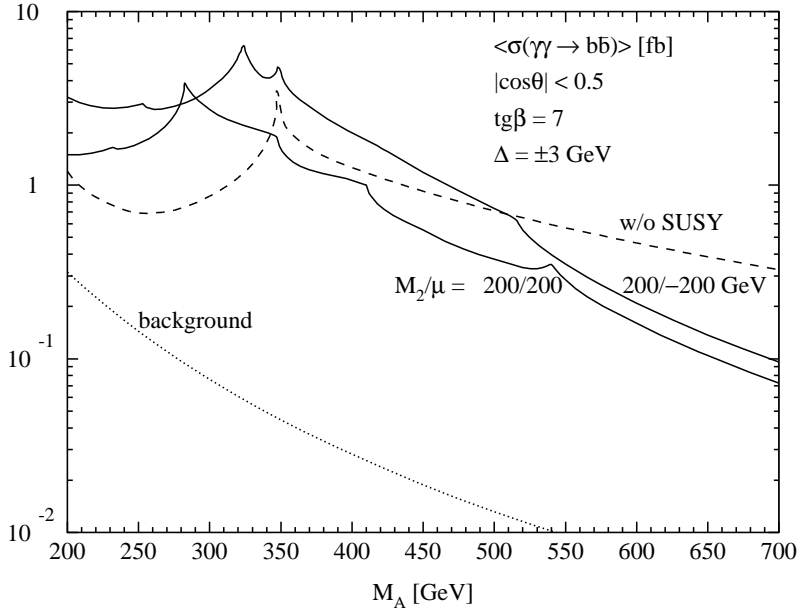,width=0.8\textwidth}
\vspace*{-0.cm}
\caption{ (a) Cross section for resonant heavy Higgs $H,A$ boson
production as a function of the pseudoscalar Higgs mass $M_A$ with
decay into $b\bar b$ pairs, and the corresponding background cross
section. The maximum of the photon luminosity has been turned to
$M_A$.  Cuts as indicated. The \MSSM\ paraameters have been chosen as
$\tan \beta =7, M_2 = \pm \mu = 200$ \GEV; the limit of vanishing
SUSY-particle contributions is shown for comparison. 
The cross sections are defined
in $b\bar b$ mass bins of $M_A\pm 3$ GeV around the maximum of the
\GG\ luminosity. See also comments in text.}
\label{fig:bot}
\end{center}
\end{figure}

The results for the cross section of the $H$, $A$ signal in the $b\bar
b$ decay mode and the corresponding background for the value of
$\tan\beta=7$ are shown in Fig.~\ref{fig:bot} as a function of the
pseudoscalar mass $M_A$~\cite{Muhlleitner}. From the figure one can
see that the background is strongly suppressed with respect to the
signal. The significance of the heavy boson signals is sufficient for
a discovery of the Higgs particles with masses up to about 70--80\% of
the $e^+e^-$ c.m.s. energy. For a 500 GeV $e^+e^-$ LC the $H,A$ bosons
with masses up to about 400 GeV can be discovered in the $b\bar b$
channel. For a LC with $\sqrt{s_{ee}}=900$ GeV
the range can be extended to about 700 GeV~\cite{4b,muehldiss}.  For
heavier Higgs masses the signal becomes too small to be detected. Note
that the cross section given in Fig.\ref{fig:bot} is related to the
luminosity $k^2 L_{geom} \sim 0.4L_{geom}$ which is $4.8\times
10^{34}$\CMS($\sim 1.5 L_{\EPEM}$) for 2E = 500 GeV and it grows
proportional to the energy.

\begin{figure*}[htb]
\vspace{3mm}
\begin{center}
\epsfig{file=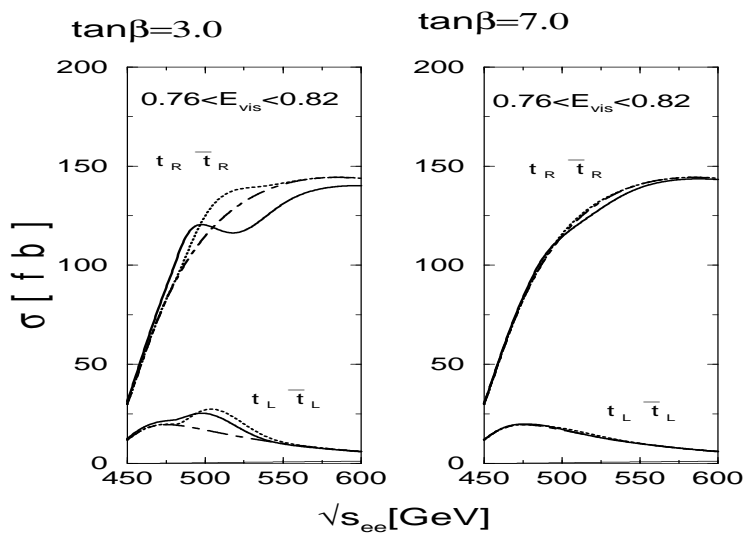,width=0.7\textwidth,height=0.5\textwidth}
\caption{The effective top pair cross sections convoluted with the
$\gamma\gamma$
  luminosity within the visible energy range as indicated.  The
  bold-solid curves correspond to the correct cross sections, the
  dotted curves are the ones neglecting the interference, and the
  dot-dashed are the continuum cross sections, respectively.  The
  upper curves are for $t_R \overline{t}_R$, and the lower ones for
  $t_L \overline{t}_L$.  The sum of the tree cross sections for $t_R
  \overline{t}_L$ and $t_L \overline{t}_R$, are also plotted in the
  thin-solid line located very near to the bottom horizontal axis.
  The left figure is for $\tan\beta$ = 3, and the right for
  $\tan\beta$ = 7~\cite{AKSW00}.  }
\label{F:AKSW2}
\end{center}\vspace{3mm}
 \end{figure*}

The almost degenerate $H$ and $A$ states can be separated by using the
linear polarization of the colliding photons (see eq.\ref{clin}). The
$H$ and $A$ states can be produced from collisions of parallel and
perpendicularly polarized incoming photons,
respectively~\cite{Yang50,GF92,KKSZ94,GK94,CK95}. Next-to-leading
order QCD corrections to the asymmetry of heavy quark production in
linearly polarized photon beams are shown to be
small~\cite{Jikia}. The possible $CP$-violating mixing of $H$ and $A$
can be distinguished from the overlap of these resonances by studying
the polarization asymmetry in the two-photon production~\cite{GIv}.

The interference between $H$ and $A$ states can be also studied in
the reaction $\GG\to t\bar{t}$ with circularly polarized photon beams by
measuring the top quark helicity~\cite{AKSW00}. The corresponding
cross sections are shown in Fig.~\ref{F:AKSW2}. The effect of the
interference is clearly visible for the value  $\tan\beta=3$.  The
$RR$ cross section is bigger than the $LL$ cross section [$R$($L$) denotes the
right(left) helicity] due to the continuum. Large interference effects
are visible in both modes.  Without the measurement of the top quark
polarization there still remains a strong interference effect between
the continuum and the Higgs amplitudes, which can be measured.

For the pair production of charged Higgses $\GG\ \to H^+H^-$, due to
much larger cross section, Fig.\ref{crossel}, the event rate at the
photon collider will be almost an order of magnitude larger than at
the $e^+e^-$ LC.

\subsection{Extended Higgs Models}

The scenario, in which all new particles are very heavy, can be
realized not only in the \MSSM\, but also in other extended models of
the \SM\ Higgs sector, for example in models with two Higgs
doublets. In this case the two-photon Higgs boson width differs from
the \SM\ value due to contributions of extra heavy charged particles,
i.e. charged Higgs bosons in the \2HDM.

Different models for the \2HDM\ have been discussed in
Ref.~\cite{GKO}. Assuming that the branching ratios of the observed
Higgs boson to quarks, $Z$ or $W$ bosons are close to their \SM\
values, two sets of possible values for the couplings are obtained.
For solution A all these couplings are close to their values in the
\SM\ up to a common sign. For solution B absolute values of the
couplings are close to their \SM\ values, but the coupling to quark
$q=u (t)$ or $d (b)$ is of opposite sign as compared to the coupling
to the gauge bosons. Fig.~\ref{fig:gaga-a,b} shows deviations of the
two-photon Higgs width from the \SM\ value for these two solutions.
The shaded regions are derived from the anticipated $1\sigma$
experimental bounds around the \SM\ values for the Higgs couplings to
fermions and gauge bosons. Comparing the numbers in these figures with
the accuracy of the two-photon Higgs width at a photon collider, it
can concluded that the difference between \SM\ and \2HDM\ should
definitely be observed~\cite{GKO}.

\begin{figure}[htb]
\begin{center}
\epsfig{file=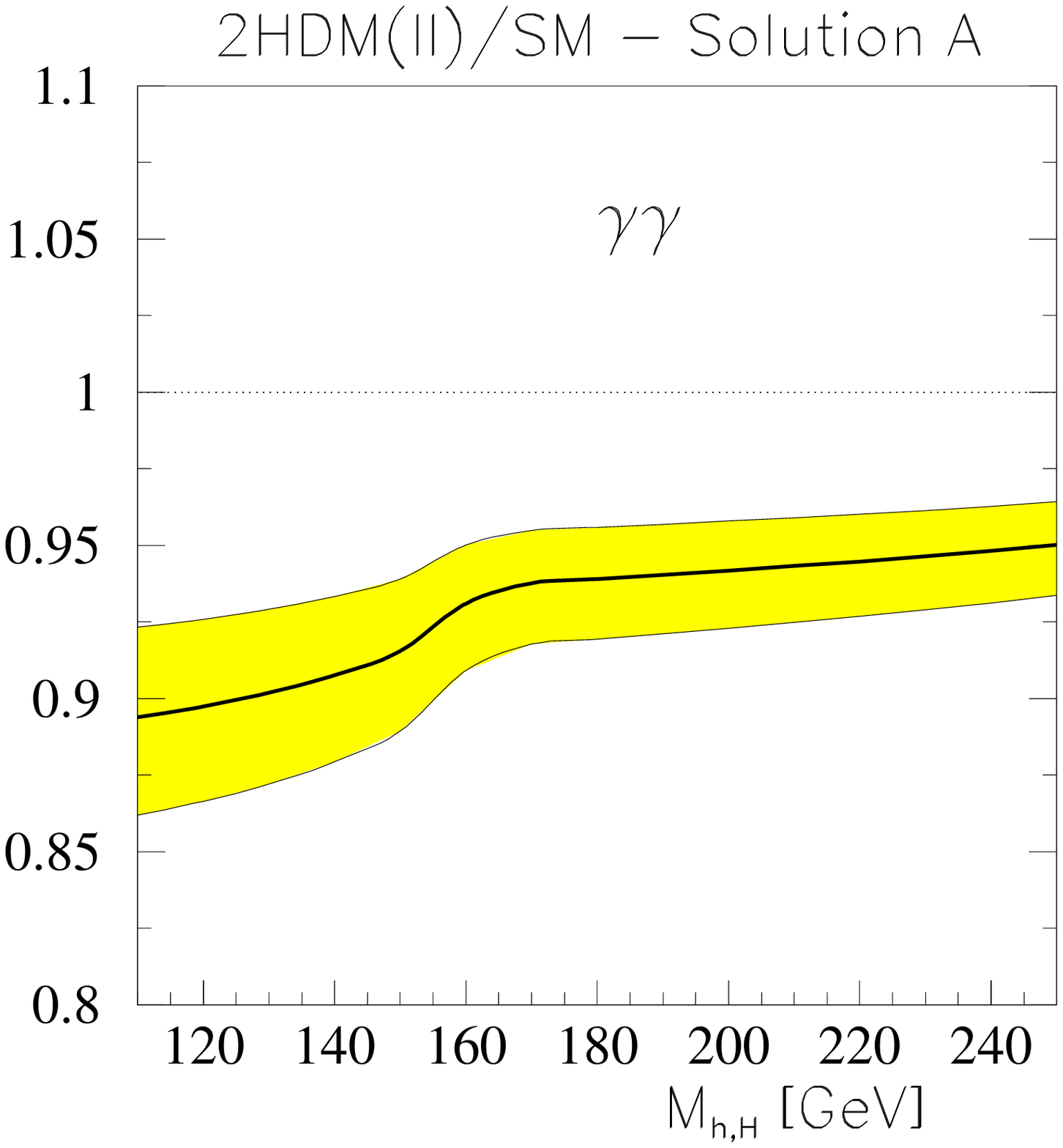,width=0.5\textwidth,height=0.4\textwidth}

\vspace*{.3cm}

\epsfig{file=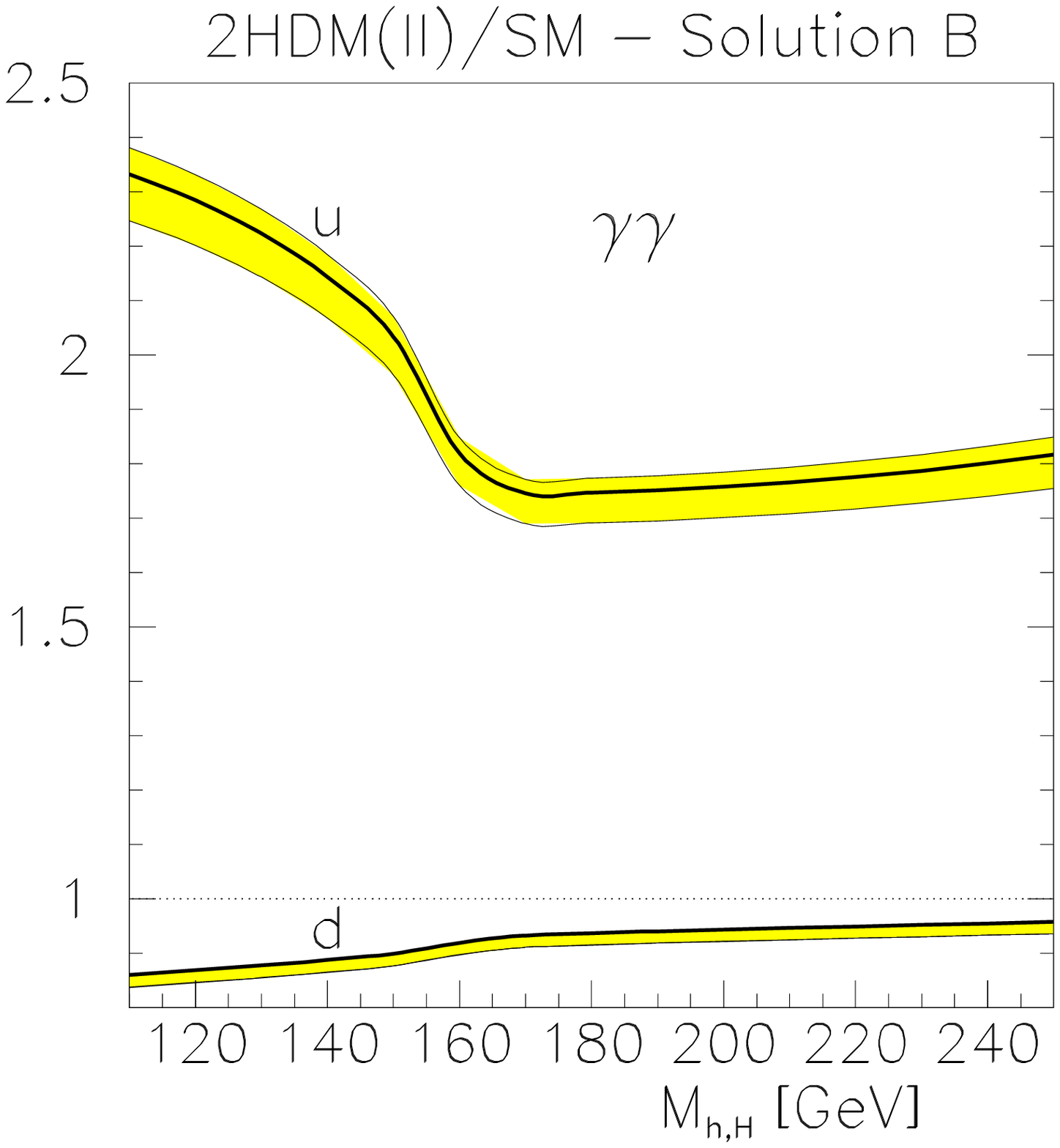,width=0.5\textwidth,height=0.4\textwidth}
\end{center}
\caption{The ratio of the two-photon Higgs width in the \2HDM to its
\SM\ value, for two different solutions~\cite{GKO}. See also the text.}
\label{fig:gaga-a,b}
\end{figure}

The $CP$ parity of the neutral Higgs boson can be measured using
linearly polarized photons.  Moreover, if the Higgs boson is a mixture
of $CP$-even and $CP$-odd states, as in a general \2HDM\ with
$CP$-violating neutral sector, the interference of these two terms
gives rise to a $CP$-violating
asymmetries~\cite{GF92,GK94,KKSZ94,ACHL00,GIv}.  Two $CP$-violating
ratios could contribute to linear order with respect to $CP$-violating
couplings:

\begin{equation} 
{\mathcal A_1}=\frac{|{\mathcal M}_{++}|^2-|{\mathcal M}_{--}|^2} {|{\mathcal
M}_{++}|^2+|{\mathcal M}_{--}|^2}, \quad {\mathcal A}_2=\frac{2Im({\mathcal
M}_{--}^*{\mathcal M}_{++})} {|{\mathcal M}_{++}|^2+|{\mathcal M}_{--}|^2}.
\end{equation}

Since the event rate for Higgs boson production in $\g\g$ collisions
is given by
\begin{eqnarray}
dN &=& dL_{\g\g}dPS \frac{1}{4}
(|{\mathcal M}_{++}|^2+|{\mathcal M}_{--}|^2) \nonumber\\
&&\times
[(1+\langle\xi_2\tilde\xi_2\rangle)
+(\langle\xi_2\rangle+\langle\tilde\xi_2\rangle){\mathcal A}_1
+(\langle\xi_3\tilde\xi_1\rangle+\langle\xi_1\tilde\xi_3\rangle){\mathcal A}_2],
\end{eqnarray}
where $\xi_i$, $\tilde\xi_i$ are the Stokes polarization parameters,
two $CP$-violating asymmetries can be observed. The  asymmetry measured with 
circularly polarized photons is given by
\begin{equation}
T_{-} = \frac{N_{++}-N_{--}}{N_{++}+N_{--}}=
\frac{\langle\xi_2\rangle+\langle\tilde\xi_2\rangle}
{1+\langle\xi_2\tilde\xi_2\rangle}{\mathcal A}_1,
\end{equation}
where $N_{\pm\pm}$ correspond to the event rates for positive
(negative) initial photon helicities. Experimentally the
asymmetry is measured by simultaneously flipping the helicities
of the laser beams used for production of polarized electrons and
$\gamma \to e$ conversion.  The asymmetry  measured with linearly
polarized photons is given by
\begin{equation}
T_\psi = \frac{N(\phi=\frac{\pi}{4})-N(\phi=-\frac{\pi}{4})}
{N(\phi=\frac{\pi}{4})+N(\phi=-\frac{\pi}{4})} = 
\frac{\langle\xi_3\tilde\xi_1\rangle+\langle\xi_1\tilde\xi_3\rangle}
{1+\langle\xi_2\tilde\xi_2\rangle}{\mathcal A}_2,
\label{poldep}
\end{equation}
where $\phi$ is the angle between the linear polarization vectors of
the photons.  The asymmetries are typically larger than 10\%
\cite{GF92,GK94,KKSZ94,ACHL00,GIv} and they are observable for a large
range of \2HDM\ parameter space if $CP$ violation is present in the
Higgs potential.

Hence, a high degree of both circular and linear polarizations for the
high energy photon beams gives additional opportunities at the \GG\
collider for the detailed study of the Higgs sector.

\section{Supersymmetry}

In $\gamma\gamma$ collisions, any kind of the charged particles can be
produced in pairs if the mass is below the kinematical bound.
Important cases for a photon collider are the charged
sfermions~\cite{JLCgg,Klasen}, the charginos~\cite{JLCgg,Mayer} and
the charged Higgs bosons.

For the \GG\ luminosities given in the Table~\ref{tabtel}, the
production rates for these particles will be larger than in \EPEM\
collisions. Hence detailed studies of charged supersymmetric particles
are possible at the \ggam\ collider.  In addition, the cross sections
in \GG\ collisions are given by pure QED to leading order, while in
\EPEM\ collisions also $Z$-boson and sometimes $t$-channel exchanges
contribute.  So, studies of these production processes in both
channels provide complementary information about the interactions of
these charged sparticles.

The \egam\ collider could be the ideal machine for the discovery of
the scalar electron and neutralino in
the reaction $e^-\gamma \to\tilde{e}^- \tilde{\chi}_1^0$ 
\cite{JLCgg,Cuyp,GK92,Blochinger}. They could be discovered
in $\gamma e$ collisions up to the kinematical limit of
\begin{equation}
M_{\tilde{e}^-} < 0.9 \sqrt{s}_{ee} - M_{\tilde{\chi}_1^0} , 
\label{eq:sebound}
\end{equation}
where $\sqrt{s_{ee}}$ is the energy of the original \EPEM\ collider.
This bound would exceed the bound  for $\tilde{e}^+ \tilde{e}^-$
pair production in  $e^+e^-$ collisions if $M_{\tilde{\chi}_1^0} <
0.4 \sqrt{s}_{ee}$.\footnote{Cross sections of the reactions
$e^+e^-\to\tilde{e}^\mp \tilde{\chi}_1^0e^\pm$ with lower threshold in
$e^+e^-$ collisions are suppressed by an extra factor of $\alpha$.}

In some scenarios of supersymmetric extensions of the Standard Model
stoponium bound states $\tilde{t}\bar{\tilde{t}}$ are formed.  A
photon collider would be the ideal machine for the discovery and study
of these new narrow strong resonances~\cite{Ilyin}.  About ten
thousand stoponium resonances for $M_{S}=200$ \GEV\ will be produced for an
integrated \GG\ luminosity in the high energy 
peak of $100$~fb$^{-1}$.  Thus precise measurements of the stoponium
effective couplings, mass and width should be possible.  At \EPEM\ 
colliders the counting rate will be much lower and in some scenarios
the stoponium cannot be detected due to the large background~\cite{Ilyin}.

\section{$W$ Boson Interactions}

One of the best known examples of new physics scenarios which clearly
show the complementarity of the $e^+e^-$ and the $\gamma\gamma$,
$\gamma e$ modes of an LC, are anomalous gauge couplings of the $W$ boson.
Recent experiments at LEP2 and the Tevatron have observed $WW$, $ZZ$
pair production and they have verified that the cross sections for the
production of weak gauge boson pairs, at least in the region close to
threshold, conform to the Standard Model predictions. New strong
interactions that might be responsible for the electroweak symmetry
breaking can affect the triple and quartic couplings of the weak
vector bosons. The precision measurements of these couplings, as well
as corresponding effects on the top quark couplings can provide clues
to the mechanism of the electroweak symmetry breaking.  Moreover, if
all new particles are very heavy, anomalous couplings are an important
source of information on the mechanism of EWSB.  Estimates
suggest that a high level of precision is needed, which is much higher than
the current bounds close to $10^{-1}$ on the parameters of the
$W$ vertices from LEP2 and the Tevatron~\cite{bounds}.  Due to the
huge cross sections, of the order of $10^2$~pb, well above the
thresholds, the $\gamma\gamma \rightarrow W^+W^-$ and $e^-\gamma
\rightarrow \nu W^-$ processes seem to be ideal reactions to study the
anomalous gauge interactions.

\subsection{Anomalous Gauge Boson Couplings}

The relevant process at the \epe collider is \eeww.  This reaction is
dominated by the large $t$-channel neutrino exchange diagram which
however can be removed using electron beam polarization. The cross
section of $W^+W^-$ pair production in $e^+e^-$ collisions with
right-handed electron beams, eliminating the neutrino exchange, has a
maximum of about 2~pb at LEP2 and decreases at higher energy.

The two main processes at the photon collider are \ggww, \gewnu.
Their total cross sections for center--of--mass energies above $200$ \GEV\
are about 80~pb and 40~pb, respectively, and they do not decrease with
energy. Hence the $W$ production cross sections at the photon collider
are at least 20--40 times larger than the cross section at the \EPEM\
collider. This enhancement makes event rates at 
the photon collider  one order of magnitude larger than at an
\EPEM\ collider, even when the lower \GG, \GE\ 
luminosities are taken into account. Specifically
for the integrated \GG\ luminosity of 100~fb$^{-1}$,  about
$8\times 10^6$ $W^+W^-$ pairs are produced at the photon collider.
Note that while  $\gamma e \to W\nu$ and $\gamma \gamma \to WW$  isolate
the anomalous photon couplings to the $W$, $\EPEM  \to WW$ involves
potentially anomalous $Z$ couplings so that the two LC modes are complementary
with each other.

While there have been extensive analyses~\cite{rev} of anomalous
triple gauge boson couplings at \EPEM\ colliders, the corresponding
processes $\gamma e\to W\nu$ and $\GG\to WW$ have generally received
less attention in the literature~\cite{old,BBB,JLCgg} and the higher
order processes were not considered in detail.

The analysis of $\GG\to WW$ has been performed in
Refs.~\cite{JLCgg,TakahashiWW} with the detector simulation. The analysis at
 photon colliders is compared to that at \EPEM\
colliders. The results have been obtained only from analyses of the
total cross section. With the $W$ decay properties taken into account
further improvements can be expected. The resulting accuracy on
$\lambda_{\gamma}$ is comparable with \EPEM\ analyses, while a similar
accuracy on $\delta \kappa_{\gamma}$ can be achieved at $1/20$--th of
the \EPEM\ luminosity.

In addition, the process  \gewnu\, which has a large cross section, is
very sensitive to the admixture of right--handed currents in the $W$
couplings with fermions:  $\sigma_{\gewnu} \propto (1-2 \lambda_e)$.

Many processes of 3rd and 4th order have quite large cross
sections~\cite{GinLCWS,3order,GinW,ginzburg} at the photon collider:

\begin{center}
\begin{tabular}{l}
$\egam\to eWW$ \\
$\egam\to \nu WZ$ \\
\end{tabular}
\begin{tabular}{c}
\qquad\qquad
\end{tabular}\begin{tabular}{l}
$  \ggam\to ZWW$    \\
$\GG\to WWWW$ \\
$\GG\to WWZZ$ \\
\end{tabular}
\end{center}

It should also be noted, that in \GG\ collisions the
anomalous  $\gamma\gamma W^+W^-$ quartic couplings can be probed.
However, the higher event rate does not
necessarily provide better bounds on anomalous couplings.  In some
models electroweak symmetry breaking leads to large deviations mainly
in longitudinal $W_L W_L$ pair production~\cite{BBB}. On the other hand
the large cross section of the reaction $\GG\to W^+W^-$ is due to
transverse $W_T W_T$ pair production. In such a case transverse
$W_T W_T$ pair production would represent a background for the
longitudinal $W_L W_L$ production.  The relative yield of $W_L W_L$ can
be considerably improved after a cut on the $W$ scattering angle.
Asymptotically for $s_{\GG} \gg  M^2_W$ the production of $W_L W_L$ is
as much  as 5 times larger than at a \EPEM\ LC. 

However, if anomalous couplings manifest themselves in transverse
$W_T W_T$ pair production, e.g. in theories with large
extra dimensions, then the interference with the large
\SM\ transverse contribution is of big advantage in the photon
collider. 

\subsection{Strong $WW\to WW$, $WW\to ZZ$ Scattering}

If a strong-coupling EWSB breaking scenario is realized in Nature, $W$
 and $Z$ bosons will interact strongly at high energy. For example, if
 no Higgs boson exists with a mass below 1~TeV, the longitudinal
 components of the electroweak gauge bosons must become strongly
 interacting at energies above 1~TeV to comply with the requirements
 of unitarity for the $W_LW_L$, $Z_LZ_L$ scattering amplitudes. In
 such scenarios novel resonances can be formed in the ${\mathcal
 O}$(1~TeV) energy range which can be produced in $W_LW_L$
 collisions. In these scenarios, $W_LW_L$ must be studied at energies
 of the order of 1~TeV. In $e^+e^-$ collisions $W_LW_L$ scattering can
 be investigated by using $W$ bosons radiated off the electron and
 positron beams in the reaction $ \EPEM\to\nu\nu W^+W^-$.  If the
 energy of the $\gamma\gamma$ collisions is high enough, the effective
 $W$ luminosity in $\gamma\gamma$ collisions becomes large enough to
 allow for the study of $W^+W^-\to W^+W^-$, $ZZ$ scattering in the
 reactions \be \g\g\to WWWW, \;\; WWZZ. \ee Each incoming photon turns
 into a virtual $WW$ pair, followed by the scattering of one $W$ from
 each such pair to form the final $WW$ or $ZZ$
 pairs~\cite{boudjema,brodsky,JikiaWWWW,CheungWWWW}.  The same
 reactions can be used to study anomalous quartic $WWWW$, $WWZZ$
 couplings.

A potential advantage of the $\gamma\gamma$ collider is the
longitudinal $W$ spectrum inside the photon~\cite{W/gamma}. Due to the
hard component, with the logarithmic enhancement factor, the
distribution function at large $z$ is bigger for photons than for
electrons. For $\sqrt{s} = 1$ TeV the $W_LW_L$ luminosity
distributions in \GG\ and \EPEM\ collisions are very similar, but
${\mathcal L}_{W_LW_L/\g\g}$ is larger than ${\mathcal
L}_{W_LW_L/e^+e^-}$ by a factor of 5~\cite{JikiaWWWW} for $\sqrt{s} =
5$ TeV.  A detailed comparison can be carried out only after accurate
simulations of the processes have been performed, taken backgrounds
into account properly.

\section{Extra Dimensions}

\begin{figure*}[htb]
\centerline{
\psfig{figure=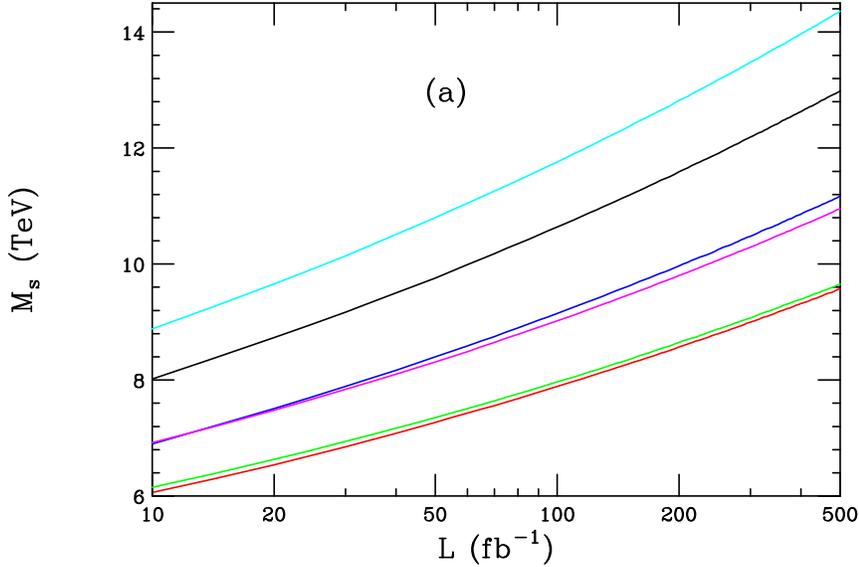,width=7.5cm,angle=90}}
\caption[*]{$M_s$ discovery reach for the process $\GG\to W^+W^-$ at a
  $2E_0=1$ \TEV\ LC  as a function of the 
  integrated luminosity for the different initial state
  polarizations assuming $\lambda=1$.  From top to bottom on the
  right hand side of the figure the polarizations are $(-++-)$,
  $(+---)$, $(++--)$, $(+-+-)$, $(+---)$, and $(++++)$.}
\label{fig:extra6}
\end{figure*}

New ideas have recently been proposed to explain the weakness of the
gravitational force~\cite{add}. The Minkowski world is extended by extra
space dimensions which are curled up at small dimensions $R$. While the gauge
and matter fields are confined in the (3+1) dimensional world, gravity
propagates through the extended 4+n dimensional world. While the effective
gravity scale, the Planck scale, in four dimensions is very large, the
fundamental Planck scale in 4+n dimensions may be as low as a few TeV so that
gravity may become strong already at energies of the present or next
generation of colliders.

Towers of Kaluza--Klein graviton excitations will be realized on the
compactified 4+n dimensional space. Exchanging these KK excitations between
\SM\ particles in high--energy scattering experiments will generate effective
contact interactions, carrying spin=2 and characterized by a scale $M_s$ of
order few TeV. They will give rise to substantial deviations from the
predictions of the Standard Model for the cross sections and angular
distributions for various beam polarizations~\cite{pheno,rizzo}.

Of the many processes examined so far, $\GG\to WW$ provides
the largest reach for $M_s$ for a given 
center of mass energy of the LC~\cite{rizzo1,rizzo}.  
The main reasons are that the $WW$ final
state offers many observables which are particularly sensitive to the
initial electron and laser polarizations and the very high
statistics due to the 80~pb cross section.

By performing a combined fit to the total cross sections and angular
distributions for various initial state polarization choices and the
polarization asymmetries, the discovery reach for $M_s$ can be
estimated as a function of the total $\GG$ integrated luminosity. This
is shown in Fig.~\ref{fig:extra6}~\cite{rizzo}. The reach is in the
range of $M_s\sim (11$--$13)\cdot 2E_0$, which is larger than that
obtained from all other processes examined so far. By comparison, a
combined analysis of the processes $\EPEM\to f\bar f$ with the same
integrated luminosity leads to a reach of only $(6$--$7)\cdot 2E_0$.

Other  final states in \GG\ collisions are also
sensitive to graviton exchanges, two examples being the
$\GG$~\cite{Cheung,Davoudiasl} and $ZZ$~\cite{rizzo1} final
states, which however result in smaller search reaches. 

\section{Studies of the Top Quark}

The top quark is heavy and up to now point-like at the same time.  The
top Yukawa coupling $\lambda_{t} = 2^{3/4}G_F^{1/2}M_{t}$ is
numerically very close to unity, and it is not clear whether this is
related to some deep physical reason.  Hence deviations from the \SM\
predictions should be expected most pronounced in the top
sector~\cite{peccei}. Studies of the top quark may shed light on the
origin of the mechanism of EWSB. Top quark physics will therefore be a
very important part of research programs for all future hadron and
lepton colliders. The $\gamma\gamma$ collider is of special interest
because of the very clean production mechanism and high rate (review
Ref.~\cite{hewett}). Moreover, the $S$ and $P$ partial waves of the
final state top quark--antiquark pair produced in $\GG$ collisions can
be separated by choosing the same or opposite helicities of the
colliding photons.

To get a reliable answer for the production cross section near the
threshold, it is necessary to resum the Coulombic corrections as for
$e^+e^-$ collisions~\cite{ee-thres}. After resummation, the cross
section close to the threshold increases by a factor
4--5~\cite{penin}. Recent results~\cite{penin} show a large difference
between the NLO and NNLO predictions.

\subsection{Probing Anomalous Couplings in $t\bar t$ Pair
Production}

Two points  are different in
$\gamma\gamma$ and $e^+e^-$ collisions with respect to the
couplings of the top quark:
\begin{itemize}
\item in $\gamma\gamma$ collisions the $\gamma t \bar{t}$ coupling 
enters with the 4th power;
\item the $\gamma t \bar{t}$ coupling is isolated in $\gamma\gamma$
collisions while in $e^+e^-$ collisions both $\gamma t \bar{t}$ and $Z
t \bar{t}$ couplings contribute.
\end{itemize}
The effective Lagrangian contains four parameters $f_i^{\alpha}$,
Ref.\cite{Boos}, where $\alpha = \gamma,Z$; but only couplings
with $\alpha = \gamma$ occur in $\gamma\gamma$ collisions.  It has
been demonstrated~\cite{djuadi} that if the cross section can be
measured with 2\% accuracy, scale parameters for new physics $\Lambda$
can be probed up to 10 TeV for $2E_0 = 500$ GeV, with the form factors
are taken in the form $f_i^{\alpha}=
(f_i^{\alpha})^{SM}(1+s/\Lambda^2)$.  The sensitivity to the anomalous
magnetic moment $f_2^{\gamma}$ is better in $\gamma\gamma$ than in
$e^+e^-$ collisions.  The $f_4^{\alpha}$ term describes the $CP$
violation. The best limit on the imaginary part of the electric dipole
moment $Im(f_4^{\gamma})$ is about $2.3\times 10^{-17}$ $e$cm~\cite{rindani}. 
It can be derived from the forward-backward asymmetry
$A_{fb}$ with initial-beam helicities of electron and laser beams
$\lambda_e^1=\lambda_e^2$ and $\lambda_l^1= -\lambda_l^2$. The
real part of the dipole moment can also be bounded to the
order  $10^{-17}$ $e$cm by using linear polarization
asymmetries~\cite{choi}. 

\begin{figure}[htbp]
\vspace{-3mm}
\centering{
\epsfig{file=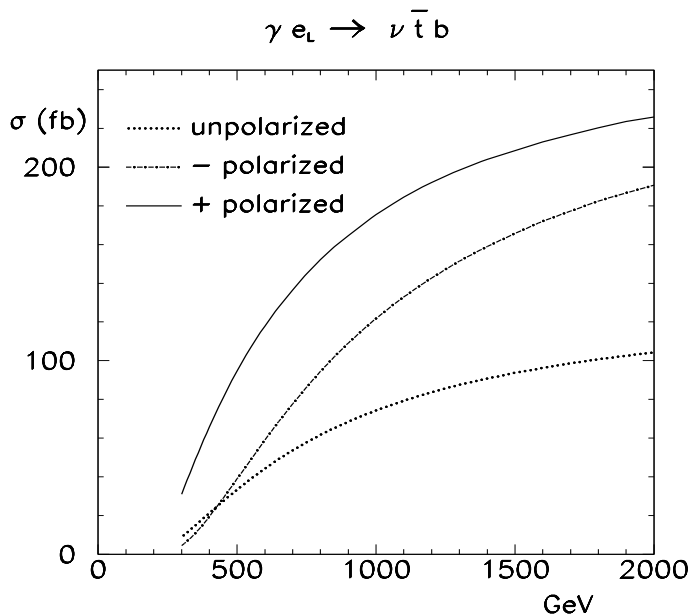,width=10cm,height=9cm}}
\caption{Single top quark production cross section in $\gamma e$
collisions}
\label{fig:f9}
\vspace{-3mm}
\end{figure}

\subsection{Single Top Production in $\gamma \gamma$ and $\gamma e$ Collisions}

Single top production in $\gamma\gamma$ collisions results in the same
final state as the top quark pair production~\cite{boos-gg} and
invariant mass cuts are required to suppress direct $t\bar t$
contribution.  Single top production can also be studied in $\gamma e$
collisions~\cite{singletop}. In contrast to the  pair production
rate the single top rate is directly proportional to the $Wtb$
coupling and the process is therefore very sensitive to its
structure. The anomalous part of the effective Lagrangian, Ref.\cite{Boos}, 
contains terms $f_{2L(R)} \propto 1/\Lambda$, where
$\Lambda$ is the scale of a new physics.

\begin{table}
\caption{Expected sensitivity for $Wtb$ anomalous couplings
measurements. The total integrated luminosity was assumed to be
500~fb$^{-1}$ for $e^+e^-$ collisions and 250~fb$^{-1}$ and
500~fb$^{-1}$ for $\gamma e$ collisions at 500~GeV and 2~TeV,
respectively.}
\vspace{0.3cm}

  \begin{center}
    \begin{tabular}{|l|c|c|}\hline
        &{$f_2^L$}
        &{$f_2^R$} \\\hline\hline
      Tevatron ($\Delta_{sys.}\approx10\%$)
               & $-0.18$ $\div$$+0.55$ & $-0.24$ $\div$$+0.25$ \\
      LHC ($\Delta_{sys.}\approx5\%$)
               & $-0.052$$\div$$+0.097$ & $-0.12$ $\div$$+0.13$ \\
$e^+e^-$ ($\sqrt{s_{e^+e^-}}=0.5$ TeV) 
               & $-0.025\div +0.025$ & $-0.2\div +0.2$ \\  
      $\gamma e$ ($\sqrt{s_{e^+e^-}}=0.5$ TeV)
               & $-0.045$ $\div$$+0.045$ & $-0.045$ $\div$$+0.045$ \\
      $\gamma e$ ($\sqrt{s_{e^+e^-}}=2.0$ TeV)
               & $-0.008$$\div$$+0.035$ & $-0.016$$\div$$+0.016$\\
      \hline
    \end{tabular}%
\vspace{5mm}
  \end{center}
\label{tb:par}
\end{table}

In the Table~\ref{tb:par}~\cite{boos-dudko-ohl,vtb-ee} limits on
anomalous couplings from measurements at different machines are
shown. The best limits can be reached at very high energy $\gamma e$
colliders, even in the case of unpolarized collisions.  In the case of
polarized collisions, the production rate increases significantly,
cf. Fig.~\ref{fig:f9}~\cite{boos-gg}, and the bounds are
improved. Only left-handed electrons lead to a nonvanishing cross
section.

\section{QCD and Hadron Physics} 

Photon colliders offer a unique possibility to probe QCD in a new
unexplored regime. The very high luminosity, the (relatively) sharp
spectrum of the backscattered laser photons and their polarization are of
great advantage.  At the photon collider the following measurements can be
performed, for example:
\begin{enumerate}
\item
  The total cross section for \GG\ fusion to hadrons~\cite{Godbole}.
\item  
  Deep inelastic \GE\ \NC\ and \CC\ scattering, and  measurement of the
  quark distributions in the photon at large $Q^2$.
\item 
  Measurement of the gluon distribution in the photon.
\item 
  Measurement of the spin dependent structure 
  function $g_1^{\gamma}(x,Q^2)$ of the photon.
\item 
  $J/\Psi$ production in \GG\ collisions as a probe of the hard QCD
  pomeron~\cite{GPS}.
\end{enumerate}

\vspace{.3cm}
{\bf \GG\ Fusion to Hadrons} 
\vspace{.3cm}

The total cross section for hadron production in \GG\ collisions is a
fundamental observable. It provides us with a picture of hadronic fluctuations
in photons of high energy which reflect the strong--interaction dynamics as
described by quarks and gluons in QCD. Since these dynamical processes involve
large distances, predictions, due to the theoretical complexity, cannot be
based yet on first principles. Instead, phenomenological models have been
developed which involve elements of ideas which have successfully been applied
to the analysis of hadron--hadron scattering, but also elements transferred
from pertubative QCD in eikonalized mini--jet models. Differences between
hadron--type models and mini--jet models are dramatic in the TESLA energy
range. \GG\ scattering experiments are therefore extremely valuable in
exploring the dynamics in complex hadronic quantum fluctuations of the
simplest gauge particle in Nature.

\vspace{.3cm}
{\bf Deep Inelastic \GE\ Scattering (DIS)}
\vspace{.3cm}

The large c.m. energy in the \GE\ system and the possibility of
precise measurement of the kinematical variables $x,Q^2$ in DIS
provide exciting opportunities at a photon collider.  In particular it
allows precise measurements of the photon structure function(s) with
much better accuracy than in the single tagged \EPEM\ collisions.  The
\GE\ collider offers a unique opportunity to probe the photon at low
values of $x$ ($ x\sim 10^{-4}$) for reasonably large values of $Q^2
\sim 10$ \GEV$^2$~\cite{PC}. At very large values of $Q^2$ the virtual
$\gamma$ exchange in deep inelastic \GE\ scattering is supplemented by
significant contributions from $Z$ exchange.  Moreover, at very large
values of $Q^2$ the charged--current process becomes effective in deep
inelastic scattering, $\GE \to \nu X$, which is mediated by virtual
$W$ exchange.  The study of this process can in particular give
information on the flavor decomposition of the quark distributions in
the photon~\cite{ZERDR}.

\vspace{0.3cm}
{\bf Gluon Distribution in the Photon}
\vspace{.3cm}

The gluon distribution in the photon can be studied in dedicated
measurements of the hadronic final state in \GG\ collisions.
The following two processes are of particular interest:
 \begin{enumerate}
\item Dijet production~\cite{gamgtojets1,gamgtojets2}, generated by
the subprocess $\gamma g \to \qqbar$;
\item Charm production~\cite{gamgtocharm}, which is sensitive to the
mechanism $\gamma g \to \ccbar$.
\end{enumerate}

Both these  processes, which are at least in certain kinematical
regions dominated by the photon--gluon fusion  mechanisms,  are sensitive to
the gluon distribution in 
the photon. The detailed discussion of these processes have been
presented in~\cite{WENGLER,ALBERT}.

\vspace{0.3cm}
{\bf Spin Dependent Structure Function
  $g_1^{\gamma}(x,Q^2)$ of the Photon} 
\vspace{.3cm}

Using polarized beams,
photon colliders offer the possibility to measure the spin dependent
structure function $g_1^{\gamma}(x,Q^2)$ of the photon~\cite{GG1}.
This  quantity is completely unknown and its measurement in
polarized \GE\ DIS would be extremely interesting for testing QCD
predictions in a broad region of $x$ and $Q^2$.  The high--energy
photon colliders allow to probe this quantity for  very small
values of $x$~\cite{JKBZGG1}.

\vspace{0.3cm}
{\bf Probing the QCD Pomeron by  $J/\Psi$ Production 
in \GG\ Collisions}
\vspace{.3cm}

The exchange of the hard QCD (or BFKL) pomeron is presumably the
dominant mechanism of the process $\GG\to J/\psi\, J/\psi$.
Theoretical estimates of the cross--section presented in~\cite{KMJPSI}
have demonstrated that measurement of the reaction $\GG\to J/\psi\,
J/\psi$ at the photon collider should be feasible.

\noindent
\begin{figure*}[htb]
\leavevmode
\begin{center}
\parbox{5.5cm}{
\epsfxsize = 5.0cm
\epsfysize = 4.2cm
\epsfbox{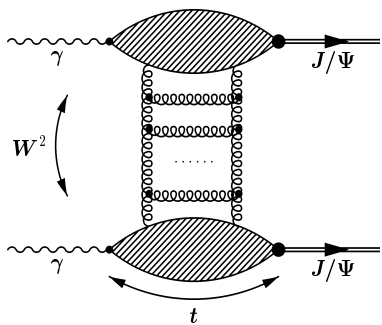}}
\end{center}
\caption{\small The QCD Pomeron exchange mechanism of the processes
         $\gamma\gamma \rightarrow J/\psi J/\psi$.}
\label{fig:qcd2}
\end{figure*}

\vspace{-6mm}

\section{Conclusions}

A short list of processes which we consider to be  most important 
for the physics program of the photon collider option of the LC, is
presented in Table~\ref{processes}.

\begin{table}[!hbtp]
\caption{Gold--plated processes at photon colliders.
\label{processes}}
\vspace{2mm}
{\renewcommand{\arraystretch}{1.2}
\begin{center}
\begin{tabular}{ l  c } 
\hline
$\quad$ {\bf Reaction} & {\bf Remarks} \\
\hline\hline
$\GG\to  h^0 \to b\bar b$ & \SM\   or \MSSM\  Higgs, 
 $M_{h^0}<160$~GeV \\
$\GG\to h^0 \to WW(WW^*)$    & \SM\ Higgs,
140\,\GEV\ $< M_{h^0}<190$~GeV \\
$\GG\to h^0 \to ZZ(ZZ^*)$      & \SM\ 
Higgs,  180\,\GEV\ $< M_{h^0}<350$~GeV \\
\hline
$\GG \to H,A\to b\bar b$  &
 \MSSM\ heavy Higgs, for intermediate $\tan\beta$\\
$\GG\to \tilde{f}\bar{\tilde{f}},\
\tilde{\chi}^+_i\tilde{\chi}^-_i,\ H^+H^-$ & large cross sections,
possible observations of FCNC \\ 
$\GG\to S[\tilde{t}\bar{\tilde{t}}]$ & 
$\tilde{t}\bar{\tilde{t}}$ stoponium  \\
$\GE \to \tilde{e}^- \tilde{\chi}_1^0$ & 
 $M_{\tilde{e}^-} < 0.9 \times 2E_0 - M_{\tilde{\chi}_1^0}$  \\
\hline
$\GG\to W^+W^-$ & anomalous $W$ interactions, extra dimensions \\
$\GE^-\to W^-\nu_{e}$ & anomalous $W$ couplings \\
$\GG\to WWWW$,$WWZZ$& strong $WW$ scatt., 
quartic anomalous $W$, $Z$  couplings\\
\hline
$\GG\to t\bar{t}$ & anomalous top quark interactions \\
$\GE^-\to \bar t b \nu_e$ & anomalous $W tb$ coupling \\
\hline
$\GG\to$ hadrons & total \GG\ cross section \\
$\GE^-\to e^- X$ and $\nu_{e}X$ & \NC\ and \CC\ structure functions
(polarized and unpolarized) \\ 
$\gamma g\to q\bar{q},\ c\bar{c}$ & gluon distribution in the photon \\
$\GG\to J/\psi\, J/\psi $ & QCD Pomeron \\
\hline
\end{tabular}
\end{center}
}

\vspace{5mm}
\end{table}

Of course there exist lots of other possible manifestations of new
physics in the $\gamma\gamma$, $\gamma e$ collisions which we have not
discussed here. The study of resonant excited electron production
$\gamma e\to e^*$, the production of excited fermions $\ggam\to f^*
f$\cite{GIVd,BBC}, leptoquark production $\egam\to
(eQ)\bar{Q}$~\cite{leptoquark}, a magnetic monopole signal in the
reaction of $\gamma\gamma$ elastic scattering~\cite{monopole} {\it
etc} may be mentioned in this context.

To summarize, the photon collider will allow us to study the physics
of EWSB in both the weak-coupling and the strong-coupling scenarios,
as well as to search for a rich spectrum of new physics
manifestations.  Measurements of the two-photon Higgs width of the
$h$, $H$ and $A$ Higgs states provide a strong motivation for
developing the technology of the $\g\g$ collider option. Photon
colliders offer unique possibilities for probing the photon structure
and the QCD Pomeron.  Polarized photon beams, large cross sections and
sufficiently large luminosities allow to enhance significantly the
discovery limits of many new particles in supersymmetric and other
extentions of the Standard Model. The accuracy of the precision
measurements of anomalous $W$ boson and top quark couplings will be
improved significantly, complementing measurements at the $e^+e^-$
mode of the linear collider.

\section*{Acknowledgements}

We greatly acknowledge our colleagues for numerous and useful
discussions on the various problems and topics. E.B. and I.G. thanks
the Organizing Committee of the Workshop for kind hospitality and
financial support. The work of E.B. was partly supported by the
RFBR-DFG 99-02-04011, RFBR 00-01-00704, and CERN-INTAS 99-377 grants.
I.G. is grateful to RFBR (grants 99-02-17211 and 00-15-96691) for
support. I.W. is supported in part by the Grant-in-Aid for Scientific
Research (No.~11640262) and the Grant-in-Aid for Scientific Research
on Priority Areas (No.~11127205) from the Ministry of Education,
Science and Culture, Japan.  J.K.  was partially supported by the EU
Fourth Framework Programme `Training and Mobility of Researchers',
Network `Quantum Chromodynamics and the Deep Structure of Elementary
Particles', contract FMRX--CT98--0194.
%
\def\MPL #1 #2 #3 {{\it Mod. Phys. Lett.} {\bf#1} (#2) #3 }
\def\NPB #1 #2 #3 {{\it Nucl. Phys.} {\bf#1} (#2) #3 }
\def\PLB #1 #2 #3 {{\it Phys. Lett.} {\bf#1} (#2) #3 }
\def\PR #1 #2 #3 {{\it Phys. Rep.} {\bf#1}  (#2) #3 }
\def\PRD #1 #2 #3 {{\it Phys. Rev.} {\bf#1}  (#2) #3 }
\def\PRL #1 #2 #3 {{\it Phys. Rev. Lett.} {\bf#1}  (#2) #3 }
\def\RMP #1 #2 #3 {{\it Rev. Mod. Phys.} {\bf#1}  (#2) #3 }
\def\NIM #1 #2 #3 {{\it Nuc. Inst. Meth.} {\bf#1}  (#2) #3 }
\def\ZPC #1 #2 #3 {{\it Z. Phys.} {\bf#1}  (#2) #3 }
\def\EJPC #1 #2 #3 {{\it E. Phys. J.} {\bf#1} (#2) #3 }
\def\IJMP #1 #2 #3 {{\it Int. J. Mod. Phys.} {\bf#1}  (#2) #3 }

\end{document}